\documentstyle[12pt]{article}

\catcode`\@=11
\def\marginnote#1{}
\newcount\hour
\newcount\minute
\newtoks\amorpm
\hour=\time\divide\hour by60
\minute=\time{\multiply\hour by60 \global\advance\minute
by-\hour}\edef\standardtime{{\ifnum\hour<12
\global\amorpm={am}%
        \else\global\amorpm={pm}\advance\hour by-12 \fi
        \ifnum\hour=0 \hour=12 \fi
        \number\hour:\ifnum\minute<10
0\fi\number\minute\the\amorpm}}
\edef\militarytime{\number\hour:\ifnum\minute<10
0\fi\number\minute}

\def\draftlabel#1{{\@bsphack\if@filesw {\let\thepage\relax
   \xdef\@gtempa{\write\@auxout{\string
      \newlabel{#1}{{\@currentlabel}{\thepage}}}}}\@gtempa
   \if@nobreak \ifvmode\nobreak\fi\fi\fi\@esphack}
        \gdef\@eqnlabel{#1}}
\def\@eqnlabel{}
\def\@vacuum{}
\def\draftmarginnote#1{\marginpar{\raggedright\scriptsize\tt#1}}
\def\draft{\oddsidemargin -.5truein
        \def\@oddfoot{\sl preliminary draft \hfil
        \rm\thepage\hfil\sl\today\quad\militarytime}
        \let\@evenfoot\@oddfoot \overfullrule 3pt
        \let\label=\draftlabel
        \let\marginnote=\draftmarginnote

\def\@eqnnum{(\theequation)\rlap{\kern\marginparsep\tt\@eqnlabel}%
\global\let\@eqnlabel\@vacuum}  }

\def\numberbysection{\@addtoreset{equation}{section}
        \def\theequation{\thesection.\arabic{equation}}}

\def\underline#1{\relax\ifmmode\@@underline#1\else
 $\@@underline{\hbox{#1}}$\relax\fi}

\catcode`@=12
\relax

\numberbysection

\advance \topmargin by -\headheight
\advance \topmargin by -\headsep

\evensidemargin \oddsidemargin
\def\rf#1{(\ref{#1})}
\def\lab#1{\label{#1}}
\def\nonu{\nonumber}
\def\br{\begin{eqnarray}}
\def\er{\end{eqnarray}}
\def\be{\begin{equation}}
\def\ee{\end{equation}}

\def\({\left(}
\def\){\right)}

\newcommand{\ct}[1]{\cite{#1}}
\newcommand{\bi}[1]{\bibitem{#1}}

\relax

\def\g{\gamma}

\def\h{{1\over 2}}

\def\o{\over}

\def\pa{\partial}
\def\pr{\prime}

\def\s{\sigma}

\def\tp0{\Theta_{+}^{(0)}}
\def\tm0{\Theta_{-}^{(0)}}

\def\vp{\varphi}

\def\ch{{\cal H}}

\def\f#1#2#3 {f^{#1#2}_{#3}}

\def\win1{{\sf w_{1+\infty}}}

\def\Win1{{\sf W_{1+\infty}}}

\def\rlx{\relax\leavevmode}
\def\inbar{\vrule height1.5ex width.4pt depth0pt}
\def\IZ{\rlx\hbox{\sf Z\kern-.4em Z}}
\def\IR{\rlx\hbox{\rm I\kern-.18em R}}
\def\IC{\rlx\hbox{\,$\inbar\kern-.3em{\rm C}$}}
\def\IN{\rlx\hbox{\rm I\kern-.18em N}}
\def\IO{\rlx\hbox{\,$\inbar\kern-.3em{\rm O}$}}
\def\IP{\rlx\hbox{\rm I\kern-.18em P}}
\def\IQ{\rlx\hbox{\,$\inbar\kern-.3em{\rm Q}$}}
\def\IF{\rlx\hbox{\rm I\kern-.18em F}}
\def\IG{\rlx\hbox{\,$\inbar\kern-.3em{\rm G}$}}
\def\IH{\rlx\hbox{\rm I\kern-.18em H}}
\def\II{\rlx\hbox{\rm I\kern-.18em I}}
\def\IK{\rlx\hbox{\rm I\kern-.18em K}}
\def\IL{\rlx\hbox{\rm I\kern-.18em L}}
\def\one{\hbox{{1}\kern-.25em\hbox{l}}}
\def\0#1{\relax\ifmmode\mathaccent"7017{#1}%
B        \else\accent23#1\relax\fi}

\def\NPB#1#2#3{{\sl Nucl. Phys.} {\bf B#1} (#2) #3}

\def\PRD#1#2#3{{\sl Phys. Rev.} {\bf D#1} (#2) #3}

\def\PLB#1#2#3{{\sl Phys. Lett.} {\bf #1B} (#2) #3}
\def\JMP#1#2#3{{\sl J. Math. Phys.} {\bf #1} (#2) #3}

\def\AoP#1#2#3{{\sl Ann. of Phys.} {\bf #1} (#2) #3}

\def\IJMPA#1#2#3{{\sl Int. J. Mod. Phys.} {\bf A#1} (#2) #3}
\def\IJMPB#1#2#3{{\sl Int. J. Mod. Phys.} {\bf B#1} (#2) #3}

\begin{document}
\begin{center}
June, 2000 \hfill{}
\vskip .4in
\hfill{}\\
{\Large {\bf The $\hat{sl}(2)$ affine Toda model coupled to the matter: solitons and confinement}}\footnote{Prepared for Hadron Physics 2000 Workshop, Caraguatatuba, SP, Brazil, 10-15 Apr, 2000.}
\end{center}

\normalsize
\vskip .4in

\begin{center}

{Harold Blas \footnotemark
\footnotetext{e-mail: blas@ift.unesp.br}}
\par \vskip .1in \noindent
Instituto de F\'{\i}sica Te\'{o}rica\\
Universidade Estadual Paulista\\
Rua Pamplona 145,\\
01405-900\,-\,S\~{a}o Paulo, S.P.\\
Brazil

\par \vskip .3in

\end{center}

\begin{center}
{\bf ABSTRACT}\\
\end{center}
\par \vskip .3in \noindent
The so-called conformal affine Toda theory coupled to the matter fields (CATM), associated to the $\hat{sl}(2)$ affine Lie algebra, is studied. The conformal symmetry is fixed by setting a connection to zero, then one defines an off-critical model, the affine Toda model coupled to the matter (ATM). The quantum version of this reduction process is discussed by means of the perturbative Lagrangian viewpoint, showing that the ATM theory is a spontaneously broken and reduced version of the CATM model. We show, using bosonization techniques that the off-critical theory decouples into a sine-Gordon model and a free scalar. Using the ``dressing'' transformation method we construct the explicit forms of the one and two-soliton classical solutions, and show that a physical bound soliton-antisoliton pair (breather) does not exist. Moreover, we verify that these solutions share some features of the sine-Gordon (massive Thirring) solitons, and satisfy the classical equivalence of topological and Noether currents in the ATM model. Imposing the Noether and topological currents equivalence as a constraint, one can show that the ATM model leads to a bag model like mechanism for the confinement of the U(1) ``color'' charge inside the sine-Gordon solitons (baryons).

\par \vskip 1.7in \noindent



\section{Introduction}

Two-dimensional field theories have been  widely explored  in the
last  years and  various phenomena  such as  dynamical mass generation,
asymptotic freedom, and quark confinement, relevant in more realistic
models, have been tested. It is well known the relevance of localized
classical solutions of non-linear relativistic field equations to the
corresponding quantum theories \ct{rajaraman}.
In particular, solitons can be associated with quantum
extended-particle states. This picture will be valid for the `gauge' field $\varphi,$ of our model. The relevance of the solitons to non-perturbative phenomena
comes, in general, from the fact that their interactions are inversely
proportional to the  coupling constants governing the dynamics of the
fundamental fields appearing in the Lagrangian. Therefore, the solitons are
weakly coupled in the strong regime of the theory, and that is the basic fact
underlying several duality ideas. The reason is that one can describe the
theory in the strong coupling regime, by replacing the fundamental Lagrangian
by another one where the excitations of its fields correspond to the solitonic
states.

We will be interested in the conformal affine Toda system coupled to the
matter fields (CATM) \ct{laf1}. In \ct{bla2} the authors have studied the classical solitonic solutions, and one of the remarkable features of the model, the equivalence of the
Noether and topological currents, and its consequences
for the confinement mechanism were outlined. The first thing to be established is if such
equivalence is not spoiled by quantum anomalies the currents may
present. Fortunately, that issue can be established exactly by using, instead
of perturbative approaches, bosonization techniques following the lines of
\ct{witten}. By bosonizing the spinor field, it is shown that the reduced theory made of the spinor and ``gauge'' particles, the so-called affine Toda model coupled to the matter (ATM), is equivalent to a theory of
a free massless scalar and a sine-Gordon field. In addition, the condition for
the equivalence of the Noether and topological currents is simply the
condition that the free massless scalar modes must be forbidden in the physical spectrum.  Therefore,
by performing a quantum reduction where the excitations of the free scalar are
eliminated, we obtain a submodel where the equivalence of those currents holds
true exactly at the quantum level, proving that there are no
anomalies. Consequently, we show that in such reduced theory the confinement
of the spinor particles does take place.  An important property of the model is that it possesses
another type of spinor particle. That is obtained by fermionizing the
sine-Gordon field using the well known equivalence between the sine-Gordon and
massive Thirring models \ct{coleman}. In that scenario the solitons
of the sine-Gordon model are interpreted as the spinor particles of the
Thirring theory. We are then lead to an interesting analogy with what one
expects to happen in QCD. The original spinor particles of our model that get
confined inside the solitons play the
role of the quarks, and the second spinor particle (Thirring), which are the
solitons, play the role of the hadrons. The $U(1)$ Noether charge is also
confined and is analogous to color in QCD (however, see a companion paper \ct{topological}). In this sense, our model constitute
an excellent laboratory to test ideas about confinement, the role of solitons
in quantum field theory, and dualities interchanging the role of solitons and
fundamental particles.

The construction of the conformal affine Toda system coupled to the matter fields (CATM),
 in the particular case of the affine Lie algebra $\hat{sl}(2),$
using the parlance of the original reference, will be summarised in the
following subsection. In section \ref{sec:dressing} we will obtain, using the dressing procedure formalism, the classical solitonic solutions of the model \ct{bla2}. In this paper we will provide the explicit forms of the two-soliton solutions and compare with the ones of the sine-Gordon theory. In section \ref{sec:quantum} we will consider some
quantum aspects of the model. We use the perturbative Lagrangian point of view to have an insight into the quantum structure of the theory \ct{bonora}. Moreover, by bosonizing the spinor field, and subsequently performing a quantum reduction, we outline a confinement mechanism present in a special class of models \ct{bla2}.

\subsection{The model}

We discuss the example associated with the principal gradation of
the untwisted affine Kac-Moody algebra $\hat{sl}(2)$. This belongs to a
special class of models introduced in \ct{laf1} possessing a $U(1)$ Noether
current depending only on the matter fields. It is then possible, under some
circunstances, to choose one solution in each orbit of the conformal group,
such that for these solutions, that $U(1)$ current is proportional to a
topological current depending only on the (gauge) zero grade fields.
The zero curvature condition in light-cone coordinates $x_{\pm
}=t\pm x$ takes the form (for $\hat{sl}(2)$ affine Lie algebra theory, notations and conventions used here, see, for example, an appendix of Ref \ct{laf2})
\br
\pa_{+} A_{-} - \pa_{-} A_{+}+\left[A_{+}\, ,\, 
A_{-}\right] =0. \lab{zerocurvature}
\er

The connections are of the form 
\begin{equation}
A_{+}=-BF^{+}B^{-1},\qquad A_{+}=-\partial _{-}BB^{-1}+F^{-}, 
\label{(4)}
\end{equation}
where the mapping $B$ is parametrized as 
\begin{equation}
B=be^{\nu C}e^{\eta Q_{{\bf s}}}=e^{\varphi H^{0}}e^{
\widehat{\nu }C}e^{\eta Q_{{\bf s}}},\,\, \mbox{with}\,\,\,\, b=e^{
\varphi \widetilde{H}^{0}},  \label{(4)}
\end{equation}
and so $\widehat{\nu }=\nu -\frac{1}{2}\varphi .$
Here $Q_{{\bf s}}$ (${\bf s}=(1,1)$) is the grading operator for the
principal gradation \ct{kac}.

The special class of models in which the $U(1)$ \ Noether current
is proportional to a topological current, occur for those models where the
first constant terms of the nonvanishing grade potentials $F^{\pm }$ are
equal to $\pm N_{{\bf s}},(N_{{\bf s}}={\bf s}_{1}+{\bf s}_{2}=2)$
respectively. So, the potentials $F^{\pm }$ are of the form
\begin{equation}
F^{+}\equiv E_{N_{{\bf s}}}+F_{1}^{+},\qquad F^{-}\equiv E_{-N_{{\bf s}%
}}+F_{1}^{-}, 
\end{equation}
where 
\br
F_{1}^{+}=2\sqrt{im}(\psi _{R}E_{+}^{0}+\widetilde{
\psi }_{R}E_{-}^{1}),F_{1}^{-}=2\sqrt{im}(
\psi _{L}E_{+}^{-1}-\widetilde{\psi }_{L}E_{-}^{0}), 
\er
and $E_{\pm N_{{\bf s}}}=mH^{\pm 1}$ ($m$ =constant)$.$

Introducing the Dirac fields $\psi ^{T}=(\psi
_{R},\psi _{L}\ )$ and substituting the explicit form of the
connections $b$ and $F_{1}^{\pm }\ \ $into \rf{zerocurvature}, we get the following equations of motion
\begin{eqnarray}
\nonu
&&\partial ^{2}\varphi\, =\,-4m_{\psi }\overline{\psi }
\gamma _{5}e^{\eta +2\varphi \gamma _{5}}
\psi ,\\
&&\partial ^{2}\widetilde{\nu }\,=\,-2m_{
\psi }\overline{\psi }(1-\gamma _{5})e^{\eta +2
\varphi \gamma _{5}}\psi -\frac{1}{2}m_{\psi
}^{2}e^{2\eta },  \nonumber \\
&&\partial ^{2}\eta \,=\,0,\,\,\,i\gamma ^{\mu
}\partial _{\mu }\psi =m_{\psi }e^{\eta +2
\varphi \gamma _{5}}\psi ,\,\,m_{\psi }\equiv 4m. 
\lab{eqn1}
\end{eqnarray}

The corresponding Lagrangian has the form
\br
\nonu
\frac{1}{k}{\cal L}=\frac{1}{4}\pa_{\mu}\vp\pa^{\mu}\vp+\frac{1}{2}\partial _{\mu }
\nu \partial ^{\mu }\eta -%
\frac{1}{8}m_{\psi }^{2}e^{2\eta }+i\overline{\psi }%
\gamma ^{\mu }\partial _{\mu }\psi -\\
m_{%
\psi }\overline{\psi }e^{\eta +2\varphi 
\gamma _{5}}\psi .   
\lab{lag1}
\er

It is real (for $\eta $ $=$real constant) if $\widetilde{
\psi }$ is proportional to the complex conjugate of $\psi ,$ and if $
\varphi $ is pure imaginary. This will be true for the particular solutions
of \rf{eqn1} such as: the 1-soliton (1-antisoliton), soliton-soliton (antisoliton-antisoliton).

The equations \rf{eqn1} are invariant under the conformal transformation
\br
 x_{+}\longrightarrow f(x_{+}),\, \, \, x_{-}\longrightarrow f(x_{-}),
\er
with $f$ and $g$ being analytic functions; and with the fields transforming
conveniently.

Therefore we can construct two chiral
currents 
\br
{\cal J}(x_{+})=m\partial _{+}\varphi +m\partial _{+}
\eta -2im\widetilde{\psi }_{R}\psi _{R}\,\,\, \mbox{and}\,\,\, 
\overline{{\cal J}}(x_{-})=m\partial _{-}\varphi +m\partial _{-}
\eta +2im\widetilde{\psi }_{L}\psi _{L},
\er
satisfying
\br
\partial _{-}{\cal J}=0,\,\,\,\,\partial _{+}
\overline{{\cal J}}=0 .
\er

Defining the current
\br
\widetilde{J}_{+}=-2im\widetilde{\psi }_{R}\psi _{R},\, \, \, \widetilde{J}_{-}=-2im\widetilde{\psi }_{L}\psi _{L},
\er
we may check that it is conserved
\br
\lab{U1}
\partial _{\mu }\widetilde{J}^{\mu }=0.  
\er

It is the $U(1)$ Noether current. In addition, this model possesses the current\br
\widetilde{j}_{+}=-m\partial _{+}\varphi , \, \,\, \widetilde{j}_{-}=m\partial _{-}\varphi ,
\er
which is a topological current, i.e., it is conserved independently of the
equations of motion
\br
\lab{topo}
\partial _{\mu }\widetilde{j}^{\mu }=0. 
\er

Under the conformal transformations the chiral currents transform as
\br
{\cal J}(x_{+})\longrightarrow (\ln f^{\prime} \left( x_{+}\right)
)^{-1}({\cal J}(x_{+})-m(\ln f^{\prime} \left( x_{+}\right) )^{\prime} ),\nonu
\\
\overline{{\cal J}}(x_{-})\longrightarrow (\ln g^{\prime}
\left( x_{-}\right) )^{-1}(\overline{{\cal J}}(x_{-})-m(\ln g^{\prime} \left(
x_{-}\right) )^{\prime} ). \lab{currentstransf}
\er

One concludes that, given a solution of the model, one can always
map it, under a conformal transformation, into a solution where
\br
\lab{reduction1}
{\cal J}(x_{+})=\overline{{\cal J}}(x_{-})=0. 
\er

Such a procedure amounts to a gauge fixing of the conformal
symmetry. We are choosing one solution in each orbit of the conformal group.
The degree of fredom eliminated corresponds to the field $\eta .$
So, in the model defined by \rf{lag1} one observes that the Noether current $
\widetilde{J}^{\mu }$ and the topological current $\widetilde{j}^{
\mu }$ are equal $\widetilde{J}_{\mu }=\widetilde{j}_{
\mu },$ as a result of the field equations.

From now on, unless otherwise stated, we consider the Lagrangian
\rf{lag1} with its conformal symmetry gauge fixed by choosing $\eta =
\eta _{0}=$const. Then we define the off-critical model
\br
\lab{lag2}
{\cal L}_{ATM}\,=\,\frac{1}{4}\partial _{\mu }\varphi \partial ^{
\mu }\varphi +i\overline{\psi }\gamma ^{\mu
}\partial _{\mu }\psi -m_{\psi }e^{\eta _{0}}%
\overline{\psi }e^{2\varphi \gamma _{5}}\psi
-\frac{1}{8}m_{\psi }^{2}e^{2\eta _{0}},
\er
which we call affine Toda model coupled to the matter (ATM). 

Now, let us study the symmetries of this Lagrangian. The
conservation law \rf{U1} corresponds to the global U(1) symmetry 
\br
\psi\rightarrow e^{i\theta }\psi ,\,\, \varphi 
\rightarrow \varphi ,\quad \widetilde{\nu }\rightarrow 
\widetilde{\nu },\,\,\mbox{with the Noether current} \ J^{\mu }=
\overline{\psi }\gamma ^{\mu }\psi .
\er

The fields $\psi $ and $\widetilde{\psi }$ have charges $1$ and $
-1$, respectively; and $\varphi $ has charge zero.

The global chiral U(1) symmetry 
\br
\lab{u1}
\psi \rightarrow e^{i
\gamma _{5}\alpha }\psi ,\qquad \varphi
\rightarrow \varphi -i\alpha ,\quad \widetilde{\nu }
\rightarrow \widetilde{\nu },
\er
also leaves the action invariant. The relevant conservation law reads 
\br
\lab{chiral1}
\partial _{\mu }[\overline{\psi }\gamma ^{\mu }\gamma _{5}
\psi +\frac{1}{2}\partial ^{\mu }\varphi ]=0.
\er

Let us next see the relationship between the Noether and
topological currents. The topological current and charge are
\begin{equation}
j^{\mu }=\frac{1}{2\pi i}\epsilon ^{\mu 
\nu }\partial _{\nu }\varphi ,\qquad Q_{topol}\equiv
\int dxj^{0}.\nonu
\end{equation}

Indeed, the Lagrangian \rf{lag1} has infinitely many degenerate vacua
due to the invariance under $\varphi \rightarrow \varphi +in
\pi .$

Combining the chiral current and the vector conservation laws$,$
and applying the arguments presented above we can set ${\cal J}$ $=\overline{%
{\cal J}}=0.$ This gives, altogether, 
\br
\frac{1}{2\pi i}\epsilon ^{\mu \nu }\partial
_{\nu }\varphi =\frac{1}{\pi }\overline{\psi 
}\gamma ^{\mu }\psi ,  \lab{equivalence}
\er
so that the topological and Noether currents are proportional.

The equation \rf{equivalence} means that the Noether density is non zero
only where $\partial \varphi \neq 0,$ that is inside the solitons.
Thus the $\psi $ field is confined inside the solitons. It will hold
at the quantum level after a suitable redefinition, we shall come back to
this point below.

\section{Dressing procedure and classical solitons}
\label{sec:dressing}

At this point the field $\psi$ seems to be a c-number field. It is well known, that in two dimensions the statistics of the
fields depends upon the coupling constant; so, we will postpone the study of
their statistics to section \ref{sec:quantum}. Field equations with the
replacement of the Fermi fields by a c-number Dirac wave function have often
been introduced, {\it ab initio}, in semiclassical calculations in the
literature, particularly in constructing models for hadrons \ct{rajaraman}.

Therefore, we will examine the classical soliton type solutions
to get insight into the quantum expectrum of the model in much the same way
as in the remarkable sine-Gordon model. We shall argue that, at the
classical level, the solutions for the $\varphi $ and $\psi $
fields share some features of the sine-Gordon and the massive Thirring
theories, respectively.

 The dressing transformation procedure provides a powerful way
of solving a wide class of nonlinear equations presenting soliton solutions
\ct{laf3}. For example, an application of the method to the vector NLS hierarchies can be found in \ct{akns}. We use this method to obtain the 1-soliton and 2-soliton solutions of the model. We want to study the behavior of the solitonic solutions taking
care of the reality conditions of the fields ($\widetilde{\psi }_{R}=
\psi _{R}^{\ast },$ $\widetilde{\psi }_{L}=\psi
_{L}^{\ast }$ and $\varphi =$ pure imaginary and $\eta =$ constant) which make the Lagrangian real.

Let us define $\varepsilon _{\pm }=mH^{\pm }$ and let $
\mid \ \widehat{\lambda }_{0}>$ \ and $\mid \ \widehat{
\lambda }_{1}>$ be the highest weight states of two fundamental
representations of the affine Kac-Moody algebra $\widehat{sl}(2),$
respectively the scalar and spinor ones. Then the solutions on the orbit of
the vaccum are the same as in the equations (10.31) of \ct{laf1}
\br
\nonu
\psi _{R}=\sqrt{\frac{m}{i}}\frac{\tau _{01}}{
\tau _{0}},\,\,\,\,\,\widetilde{\psi }_{R}=-\sqrt{
\frac{m}{i}}\frac{\tau _{12}}{\tau _{1}},\\
\lab{fields}
\psi _{L}=-\sqrt{\frac{m}{i}}\frac{\tau _{11}}{\tau 
_{1}},\,\,\,\,\,\,\widetilde{\psi }_{L}=-\sqrt{\frac{m}{i}}\frac{
\tau _{02}}{\tau _{0}},\\
\nonu
e^{-\varphi }=\frac{
\tau _{1}}{\tau _{0}},\,\,\,\,\,\,e^{-(\widetilde{\nu }-
\nu _{o})}=\tau _{0}, 
\er
where the so-called {\bf tau functions} are given by
\br
\tau _{0,1}=<\widehat{\lambda }_{0,1}\mid& G&\mid \ \widehat{
\lambda }_{0,1}>,\\
\tau _{01}=<\widehat{
\lambda }_{0}\mid E_{-}^{1}G\mid \ \widehat{\lambda }_{0}>&,&
\tau _{02}=<\widehat{\lambda }_{0}\mid GE_{+}^{-1}\mid \ 
\widehat{\lambda }_{0}>,\\
\tau _{12}=<\widehat{\lambda }_{1}\mid E_{+}^{0}G\mid \ 
\widehat{\lambda }_{1}>&,&\tau _{11}=<\widehat{
\lambda }_{1}\mid GE_{-}^{0}\mid \ \widehat{\lambda }_{1}>, 
\er
with \ G=e$^{x_{+}\varepsilon _{+}}e
^{-x_{-}\varepsilon _{-}}\,\rho\,\, e^{x_{-}\varepsilon
_{-}} e^{-x_{+}\varepsilon _{+}}$; $\rho$ being a constant group element of the $\widehat{SL}(2)$ Kac-Moody group.

\subsection{The 1-soliton solutions}

The soliton solutions of the system are constructed as
follows. Let us choose 
\br
\rho \,=\,e^{\sqrt{i}\,a_{+}V_{+}(z)}\, e^{
\sqrt{i}\,a_{-}V_{\_}(-z)},
\er
with \ [$\varepsilon _{\pm }$,V$
_{\pm }$]=$\varpi _{\pm }^{\pm }$V$_{\pm }$. The particular  factor 
$\sqrt{i}$ is chosen such that the reality condition will be obeyed
with $a_{+}=a_{-}^{\ast }$ .

Computing the matrix elements one gets the explicit form of the
solutions generated by $\rho $
\br
&&\vp = 2 \arctan \( \exp \( 2m_\psi \, \( x-x_0-vt\)/\sqrt{1-v^2}\)\), \nonu\\
\nonu
&&\psi = e^{i\theta} \sqrt{m_\psi} \, 
e^{m_\psi \( x-x_0-vt\)/\sqrt{1-v^2}}\,\\
&& \(
\begin{array}{c}
 \left( { 1-v\o 1+v}\right)^{1/4}  
{1 \o 1 +  ie^{2 m_\psi \( x-x_0-vt\)/\sqrt{1-v^2}}}\\
- \left( { 1+v\o 1-v}\right)^{1/4}  
{1 \o 1 - ie^{2 m_\psi \( x-x_0-vt\)/\sqrt{1-v^2}}} 
\end{array}\), 
\nonu\\
&&\nu = 
 \h \log \( 1 + \exp \( 4 m_\psi  \( x-x_0-vt\)/\sqrt{1-v^2}\)\)\nonu 
\\
&&+ {1\o 8} m_{\psi}^2 x_{+} x_{-},
\nonu\\
&&\eta  =  0, 
\lab{solsimple}
\er
and $\widetilde{\psi }$ is the complex conjugate of $\psi .$
The $\psi$ field solution is similar to the 1-soliton solution of the massive Thirring model \ct{orfanidis}. Notice that the $\vp$ solution is of the sine-Gordon type soliton/antisoliton \ct{rajaraman}. 

It can be directly verified that these solutions satisfy the\
relation \rf{equivalence}. It is clear \ from
the explicit expressions for the fermion field and the scalar,\ that $
\psi $ vanishes exponentially when $x-x_{0}\longrightarrow \pm \infty
,$ so that the Dirac field \ is confined inside the soliton.

\subsection{The 2-soliton solutions}

Now let us move to the study of the 2-soliton solutions of the
model. As usual in this case consider the following steps:

Let us choose \br
G=e^{x_{+}\varepsilon _{+}}e^{-x_{-}
\varepsilon _{-}}\,\rho\,\, e^{x_{-}\varepsilon
_{-}}e^{-x_{-}\varepsilon _{+}},
\er
and as the constant group element 
\br
\rho =e^{b_{1}V_{+}(\rho
_{1})}e^{a_{1}V_{-}(\nu _{1})}e^{b_{2}V_{+}(\rho
_{2})}e^{a_{2}V_{-}(\nu _{2})}.
\er

Defining $A_{i}=a_{i}e^{-\Gamma (\nu _{i})},$ $
B_{i}=b_{i}e^{\Gamma (\rho _{i})}$ and\ $\Gamma (z)=2m(zx_{+}-\frac{1
}{z}x_{-}),$\ and using the properties of level 1 vertex operators  to
determine the matrix elements in the {\sl tau functions} \ct{kac, akns}, one can obtain the {\bf soliton-soliton}
and the {\bf soliton-antisoliton} solutions.

By direct substitution, one can verify that these solutions,
indeed, satisfy the equation \rf{equivalence} without any further restriction. This allows us to conclude that, also, for 2-soliton type solutions the
Noether density is non zero only where $\partial \protect\varphi \neq 0,$
that is inside the solitons.

Let us choose the parameters conveniently in order to express in
the form of a {\bf soliton-soliton}  type solution.

Choosing the following relationships between the parameters $\nu _{1+}\nu
_{2}+\nu _{1}^{-1}+\nu _{2}^{-1}=0,$ \ $\nu _{1}
\nu _{2}=-1,$  and defining $\ u=(\nu _{1}^{2}-1)/(
\nu _{1}^{2}+1),$ one can write 
\br
\lab{solsol}
\varphi _{SS}/2i=\arctan \left(
8u^{3}\sqrt{1-u^{2}}\frac{\sinh (8m\gamma x-\log \overline{a}
)}{\cosh (8m\gamma ut)-2u\sqrt{1-u^{2}}}\right) , 
\er
where $\gamma =1/\sqrt{1-u^{2}},$  $u<1.$

Its asymptotic behaviour in time can be written as 
 \br
\frac{\varphi _{SS}}{2i}({t\rightarrow -\infty})
\Longrightarrow \tan^{-1} (e^{8m\gamma (x-u(t+\frac{\delta 
_{SS}}{2}))})-\tan^{-1} (e^{-8m\gamma (x+u(t+\frac{\delta _{SS}
}{2}))}), 
\er
\br
\frac{\varphi _{SS}}{2i}({t\rightarrow +\infty})
\Longrightarrow \tan^{-1} (e^{8m\gamma (x+u(t-\frac{\delta 
_{SS}}{2}))})-\tan^{-1} (e^{-8m\gamma (x-u(t-\frac{\delta _{SS}
}{2}))}),\\
\nonu
\delta _{SS}=\frac{1}{4mu
\gamma }\log (8u^{2}\sqrt{1-u^{2}}). 
\er

Therefore, this solution correponds to two solitons approaching one another
with relative velocity $u$ in the distant past. Notice that the time delay $
\delta _{SS},$ of course, differs from the corresponding
soliton-soliton interaction in the sine-Gordon model \ct{rajaraman}. We will come back
to this point below, when we compare our soliton solutions with the
corresponding solitons of the sine-Gordon theory.

In order to obtain {\bf soliton-antisoliton} type solution we convert
the real parameters into an imaginary ones by making $\nu 
_{1}\longrightarrow i\nu _{1},$ \ \ \ $\nu 
_{2}\longrightarrow -i\nu _{2}$\ ,\ $m\longrightarrow im.$

Setting the relationships $\ -\nu _{1}+
\nu _{2}-\nu _{1}^{-1}+\nu _{2}^{-1}=0,$ \ \ \ \ $
\nu _{1}\nu _{2}=1,$
we may obtain 
\br
\lab{solantisol}
\varphi _{S\overline{S}}/2i=\arctan \left(-
\frac{16}{u^{2}}\frac{\sinh (8m\gamma ut+\log \overline{a})}{\cosh
(8m\gamma x-\log (4/\gamma u^{2}))-2/\gamma u^{2}}%
\right) , 
\er
with $\gamma =1/\sqrt{u^{2}-1},$ \ \ $u>1$.

Even though the parameter complexifications do not modify the
purely imaginary character of $\varphi ,$ the reality condition $
\widetilde{\psi }=\psi ^{\ast }$ of the $\psi $
field is lost, converting the Lagrangian into a complex one$.$ Moreover,
with this choice of parameters the velocity $u$ becomes greater than the
velocity of light, and the mass of the soliton becomes imaginary. Nevertheless, from the mathematical point of view, still we
can write the asymptotic behaviour, extracting a  `soliton' and an `antisoliton' approaching one another with relative velocity $u$ in the
distant past and `time' delay $\delta =\frac{1}{2m\gamma }\log (u/4).$

Therefore, we have not obtained a physical bound soliton-antisoliton pair (breather) solution of the model. This fact is quite remarkable if one intends to study the quantum spectrum of the model. A discussion of this point is presented in a companion work \ct{topological}.  

In order to have a better insight into the behaviour of the two-solitons of our model, let us consider some known facts about the sine-Gordon field $\phi $ solitons. The $\phi$ field is related to its
corresponding tau functions as $-i\phi =\ln (\frac{\tau _{+}
}{\tau _{-}}).$ The tau functions of the N-soliton solutions of the
sine-Gordon equation are given by \ct{babelon}
\begin{equation}
\tau _{\pm }^{(N)}(z_{+},z_{-})=\det (1\pm V), 
\end{equation}
with $V$ a $N$ x $N$ \ matrix with elements 
\br
V_{ij}=2\frac{\sqrt{\mu _{i}\mu _{j}}}{\mu 
_{i}+\mu _{j}}\sqrt{X_{i}X_{j}},\,\,\,\, \mbox{where}\,\,\,
X_{i}=a_{i}e^{2(\mu _{i}z_{+}+\frac{1}{\mu _{i}}z_{-})},\,\,
z_{\pm}=2mx_{\pm}. 
\er

For the 2-soliton solution the determinant takes a simple form.
With a convenient choice of the parameters, one can write for the {\bf %
soliton-soliton} and {\bf soliton-antisoliton} 
\begin{equation}
\phi _{SS}/2=\tan^{-1} \left( u\frac{\sinh (8m\gamma x)}{\cosh
(8m\gamma ut)}\right) ,\,\,\,\phi _{S\overline{S}
}/2=\tan^{-1} \left( \frac{1}{u}\frac{\sinh (8m\gamma ut)}{\cosh (8m
\gamma x)}\right) ,  
\end{equation}
with $\gamma =1/\sqrt{1-u^{2}},\ \ \ u<1.$

If one makes the change $u\rightarrow iv$ in $\phi _{S
\overline{S}}$ one gets 
\begin{equation}
\phi _{v}/2=\arctan \left( \frac{1}{v}\frac{\sinh (16m
\gamma vt)}{\cosh (16m\gamma x)}\right) ,  
\end{equation}
which still gives a solution of the sine-Gordon equation. It is the
doublet or breather solutions, which represents a soliton and antisoliton
oscillating about a common center.

Then we can conclude that our solution \rf{solsol} resembles the
sine-Gordon soliton-soliton, at least asymptotically; the differences manifest in their time delays, which can be interpreted as the effect of the spinor field $\psi$ on the Toda field $\varphi$, since our model couples
those fields. Also, as in the sine-Gordon case, the equation \rf{solantisol} can be put asymptotically in the form of the mathematical `soliton' and `antisoliton', even though, each of them with imaginary mass. Notice that in
the soliton-soliton sector the time delay $\delta _{SS}$ can
eventually become negative, unlike the corresponding $\delta $ of
the sine-Gordon model, which is always positive, indicating in the latter case a repulsive force \ct{rajaraman}.

To conclude this section, let us point out that the model possesses at the classical level some solitonic
solutions which resemble the sine-Gordon and the massive Thirring models,
for the scalar and the Dirac fields, respectively. We have verified, up to
the 2-soliton solution, the equivalence between the Noether and topological
currents. The classical breather solution was not found, and it is expected that it will not appear at the quantum level.

\section{ Quantum aspects of the model}
\label{sec:quantum}

The aim of the present section is to study some quantum aspects of the model
defined by the Lagrangian \rf{lag1}.

\subsection{Perturbative Lagrangian viewpoint and the CATM $\rightarrow$ ATM reduction }
\label{subsec:perturbative}

In this subsection we will be interested in verifying how
the classical reduction of the CATM to the ATM \ct{bla1} by setting $\eta =\eta _{0}=$constant is recovered at the quantum level. This will give us some information about the vacuum of the theory, in particular about the trivial vacuum (non-topological) of the ATM model. We use the perturbative Lagrangian point of view to have an insight into the quantum structure of the theory. Let us begin expanding the action around a simple solution of the classical equations of motion \rf{eqn1}
\br
\lab{class}
\vp\,=\,\eta\,=\,\psi\,=\widetilde{\psi}\,=0,\,\,\,\nu\,=\,-\frac{1}{8}m_{\psi}^{2}x_{+}x_{-}.  
\er   

Denote collectively all the fields as
\br
\Phi=\Phi_{0}+ \Phi^{\prime},
\er
where $\Phi_{0}$ represent the relevant classical solutions \rf{class}. Let us expand every field around \rf{class}. Then relabelling the $\Phi^{\prime}$ with the same symbol as the original ones, the action corresponding to \rf{lag1} turns into
\br
\nonu
S &=& \int d^{2}x\,\Big\{ \frac{1}{4} \vp \Box \varphi -\frac{1}{2}\nu \Box \eta-\frac{1}{8} m_{\psi}^{2}-\frac{1}{4} m_{\psi}^{2}+i\overline{\psi}\g_{\mu}\pa^{\mu}\psi-m_{\psi}\overline{\psi}\psi-2im_{\psi}\overline{\psi}\g_{5}\psi\varphi -\\&&m_{\psi}\eta \overline{\psi}\psi+...\Big\}, 
\er
where the ellipsis denotes quartic or higher interaction terms in which the $\nu$ field never appears, and $\Box=\pa_{t}^{2}-\pa_{x}^{2}$. Next let us perform a Wick rotation and compute the Fourier transformed propagators. They become 
\br
<\vp,\vp>&=&\frac{1}{\h k^{2}}, \,\,\,\,\,\,<\eta,\nu>\,=\,\frac{1}{-\h k^{2}},\\
<\nu,\nu>&=&\frac{\h m_{\psi}^{2}}{\frac{1}{4} k^{4}},\,\,\,<\psi,\overline{\psi}>\,=\,\frac{\g_{\mu}p^{\mu}+im_{\psi}}{p^{2}+m_{\psi}^{2}},
\er
where $k^2=k_{0}^2+k_{x}^2$. All the other propagators vanish. Let us observe that the pole at $p^2 = - m_{\psi}^2$ in the spinor field propagator, corresponds to a massive particle of the ATM model. Let us inspect now a scattering process whose external legs consist only of the $\psi\, (\overline{\psi})$ and $\vp$ particles. From the structure of the propagators, and of the interaction terms it is easy to see that no connected graph relevant for this process will have propagation of other modes than $\psi\,(\overline{\psi})$ and $\vp$ themselves. That is to say, the other modes decouple. This fact allows us to set the fields $\nu$ and $\eta$ to zero in the action. Therefore, the action becomes exactly the action of the ATM model. In other words, for scattering processes involving massless $\vp$ modes and massive $\psi$ modes, the theory may be described by the perturbative ATM model.  

Then we have derived an effective Lagrangian with $\nu=0$, $\eta =\eta
_{0}=$const., this is exactly the off-critical affine Toda model coupled to the matter (ATM) defined above, Eq. \rf{lag2}. Then we can regard the ATM model as a spontaneously broken and reduced version of the CATM.

The symplectic structure of the reduced model \rf{lag2} has recently been studied \ct{bla1}. It was performed in the context of Faddeev-Jackiw and (constrained) symplectic methods (Wotzasek, Montani and Barcelos-Neto); by imposing the equivalence between the Noether and topological currents, Eq. \rf{equivalence}, as a constraint, the authors have been able to obtain either, the sine-Gordon model or the massive Thirring model, through a Hamiltonian reduction and gauge fixing the symmetries of the model in two different ways.

\subsection{Bosonization procedure and the ATM model}

In two-dimensional quantum field theories it is possible to transform Fermi fields into Bose fields,
and vice versa (for a complete review of the most important references in
the field see, e.g., the second Ref. of \ct{coleman}). The existence of such a
transformation, called bosonization, provided a powerful tool to obtain
nonperturbative information in two-dimensional field theories.

Let us consider the ATM Lagrangian \rf{lag2} in a slightly modified form. This modification is undertaken by imposing a convenient reality conditions (e.g. $\widetilde{\psi}=-\psi^{*}$, $\vp$ pure imaginary) on the fields and dropping a overall minus sign. See Refs. \ct{topological, bla1} for discussions concerning the positive-definite caracter of the kinetic terms of the real Lagrangian submodel, and \ct{bla2} for the implications of these conditions on the solitonic solutions of the original model. Then let us write the real action in a form which is more
convenient for the bosonization of the fermion bilinears
\br
\nonu
&&\frac{1}{k}S \,=\,\int d^{2}x\{\frac{1}{4}\partial _{\mu
}\varphi \partial ^{\mu }\varphi +i\overline{
\psi }\gamma ^{\mu }\partial _{\mu }\psi -\\
&&m_{
\psi }e^{\eta _{0}}[\overline{\psi }\frac{(1+
\gamma _{5})}{2}\protect\psi e^{2i\protect\varphi }+\overline{\psi }%
\frac{(1-\gamma _{5})}{2}\psi e^{-2i\varphi }]
-\frac{1}{8}m_{\psi }^{2}e^{2\eta _{0}}\},  \lab{action1}
\er
where we have considered the pure imaginary character of the scalar field by
making the change $\varphi \rightarrow i\varphi $ in \rf{lag2}.

The model \rf{action1} was considered in Refs. \ct{bla2, witten}. Some of the points that folllow were discussed in those papers. The model defined by \rf{action1} possesses a chiral symmetry $
\psi \rightarrow e^{i\beta\gamma _{5}}\psi ,$ $
\varphi \rightarrow \varphi -\beta $. Apparently, if
unbroken, the symmetry would prevent the fermions from having a mass. As we
will see, however, the symmetry is not broken, but the fermion has a mass.

Following \ct{coleman} one can use the boson representation of fermions as 
\begin{equation}
i\overline{\psi }\gamma ^{\mu }\partial _{\mu
}\psi =\frac{1}{2}(\partial _{\mu }c)^{2}, 
\end{equation}
\begin{equation}
\overline{\psi }(1\pm \gamma _{5})\psi =\mu \exp (\pm i
\sqrt{4\pi }c),  
\end{equation}
\begin{equation}
\overline{\psi }\gamma ^{\mu }\psi =-\frac{1
}{\sqrt{\pi }}\epsilon ^{\mu \nu
}\partial _{\nu }c.
\end{equation}

Then the action now becomes 
\begin{eqnarray}
S_{B}&=&\int d^{2}x\{\frac{1}{2}\partial _{\mu
}\varphi \partial ^{\mu }\varphi +\frac{1}{2}(\partial _{\mu }\phi)^{2}
\nonu
\\
&&-\frac{1}{2}\mu m_{\psi
}[\exp i[\sqrt{8/k}\,\varphi +\sqrt{4\pi }\phi]+\exp -i[\sqrt{8/k}\,\varphi +\sqrt{4\pi }
\phi]]\}.  
\end{eqnarray}

Introducing new fields 
\begin{equation}
\widetilde{c}\equiv \frac{\varphi/a +\sqrt{4\pi }\phi}{\sqrt{4\pi +1/a^{2}}},\qquad \widetilde{
\sigma }\equiv \frac{-\phi/a+\vp \sqrt{4\pi }}{\sqrt{4\pi+1/a^{2}}},
\end{equation}
the action takes the form 
\begin{equation}
\lab{bosonized}
S_{sG+\widetilde{\sigma}}=\int d^{2}x[\frac{1}{2}(\partial _{\mu }\widetilde{c})^{2}
+\frac{1}{2}(\partial _{\mu }\widetilde{\sigma })^{2}+\mu m_{\psi }\cos ( \sqrt{4\pi+1/a^{2}}\, \widetilde{c}),
\end{equation}
where $a^{2}\equiv k/8$.  

Therefore we obtain a
sine-Gordon model for the field $\widetilde{c}$ and a free, massless scalar $\widetilde{\sigma }$ field.

Why do the fermion and antifermion acquire masses
despite chiral symmetry? To answer this issue, we should ask what form the
original chiral current takes in terms of $\widetilde{c}$ and $\widetilde{
\sigma }$. The chiral current defined in section 1 is 
\begin{equation}
A_{\protect\mu }\equiv i\overline{\protect\psi }\protect\gamma _{\protect\mu
}\protect\gamma _{5}\protect\psi +\frac{i}{2}\partial _{\protect\mu }\protect%
\varphi .    \label{68}
\end{equation}

In terms of the new variables, one finds
\begin{equation}
A_{\protect\mu }=\frac{i}{2}\partial _{\protect\mu }\widetilde{\protect%
\sigma }.    \label{69}
\end{equation}

Thus the chiral current involves only $\widetilde{\sigma }$ and not $
\widetilde{c}.$ It is conserved at the quantum level. This means that the
field $\widetilde{c}$, and therefore also the physical fermion and
antifermion associated with this field, are neutral under chirality. Thus,
even though the elementary fermion field $\psi $ has nonzero
chirality, the physical fermion particle has zero chirality.

The presence of the physical fermions (zero chirality particles)
can be clarified introducing a new fermion field that has the quantum
numbers of the physical particles. We simply introduce a new fermion field $
\widehat{\psi }$ with 
\br
-\frac{1}{2\pi \alpha }\epsilon ^{\mu 
\nu }\partial _{\nu }\widetilde{c}=\overline{\widehat{%
\psi }}\gamma ^{\mu }\widehat{\psi }. 
\er

According to the standard rules which identify the sine-Gordon theory to the
charge-zero sector of the massive Thirring model \ct{coleman, naon}, the action\rf{bosonized} can now be written as
\br
S=\int d^{2}x[i\overline{\widehat{\psi }}
\gamma ^{\mu }\partial _{\mu }\widehat{\psi }
-m_{F}\overline{\widehat{\psi }}\widehat{\psi }-\frac{1}{2}g(
\overline{\widehat{\psi }}\gamma ^{\mu }\widehat{
\psi })^{2}+\frac{1}{2}(\partial _{\mu }\widetilde{
\sigma })^{2}].    \lab{Th+free}
\er
where 
\br
\lab{relation}
\frac{4\pi +1/a^2}{4\pi}=\frac{1}{1+g/\pi}. 
\er

Then, we observe that our theory consists of a massive fermion
with self-interaction, and a free, massless scalar.

We now discuss the quantum version of the reduction \rf{reduction1}.  
The chiral currents become 
\br
{\cal J} = i/2 \,\,\pa_{+} \widetilde{\s}  
\; , \qquad 
{\bar {\cal J}}= i/2 \,\,\pa_{-} \widetilde{\s}.
\er

Therefore, one can argue that semiclassically, the constraints \rf{reduction1} are equivalent to 
$\widetilde{\s}= {\rm const.}$, and consequently the degree of freedom eliminated by
them corresponds to the $\widetilde{\s}$ field. 
  
The reduction at the quantum level can be realized as follows. Since the
scalar field $\widetilde{\s}$ is decoupled from all others fields in the theory
\rf{bosonized}, we can denote the  space of states as 
$\ch = \ch_{\widetilde{\s}}\otimes \ch_0$, 
where $\ch_{\widetilde{\s}}$ is the Fock space of the free massless scalar
$\widetilde{\s}$, and $\ch_0$ carries the states of the rest of the theory. We shall
denote $\widetilde{\s} = \widetilde{\s}^{(+)} + \widetilde{\s}^{(-)}$, where $\widetilde{\s}^{(+)}$ ($\widetilde{\s}^{(-)}$)
corresponds to the 
part containing annihilation (creation) operators in its expansion on plane
waves.

The reduction corresponds to the  restriction of the theory to those states
satisfying  
\be
\pa_{\pm} \widetilde{\s}^{(+)} \mid \Psi \rangle = 0. 
\lab{reduce}
\ee

By taking the complex conjugate one gets
\be
\langle \Psi \mid  \pa_{\pm} \widetilde{\s}^{(-)} = 0,
\lab{reducea}
\ee
and consequently the expectation value of $\pa_{\pm} \widetilde{\s} $ vanishes on such
states. Indeed, if $\mid \Psi \rangle $ and $\mid \Psi^{\pr} \rangle $ are two
states satisfying \rf{reduce}, then 
\be
\langle \Psi^{\pr} \mid \pa_{\pm} \widetilde{\s} \mid \Psi \rangle  = 
\langle \Psi^{\pr} \mid[ \pa_{\pm} \widetilde{\s}^{(-)} + \pa_{\pm} \widetilde{\s}^{(+)}] \mid \Psi
\rangle  
= 0. 
\lab{reduceb}
\ee

That provides the correspondence with the classical constraints
\rf{reduction1}. Therefore, the Hilbert space of the reduced theory is
$\ch_{c} = \mid \Psi  \rangle \otimes \ch_0$. 

For the theory described by $\ch_{c}$, the equivalence between Noether and
topological currents, given by \rf{equivalence}, holds true at the quantum
level, since as we have shown before, \rf{equivalence} is equivalent to the
vanishing of the currents ${\cal J}$ and ${\bar {\cal J}}$. Notice that such
quantum equivalence is exact, since we have not used 
perturbative or semiclassical methods.  

One of the consequences of that quantum equivalence is that in the states of 
$\ch_{c}$, like the one-soliton \rf{solsimple}, where the space derivative
of $\vp$ is localized, one has that the spinor
$\psi$ is confined, since from \rf{equivalence}, $\langle \pa_x \vp \rangle
\sim \langle \psi^{\dagger} \psi \rangle$. Therefore, we have shown that the
confinement of $\psi$ does take place in the quantum  theory. 

The properties of the theory \rf{action1} at the quantum level are quite
remarkable. In the weak coupling regime, i.e. small $k$, the excitations around
the vacuum correspond to the spinor $\psi$ and the ``gauge''particle
$\vp$. The $U(1)$ symmetry \rf{u1} is not broken and the charged states
correspond to the $\psi$ particles. Consider now those states satisfying
\rf{reduce} and look for the fluctuations around the state corresponding, for
instance, to the one-soliton solution \rf{solsimple}. The $\psi$
particles disappear from the spectrum since they are confined inside the
soliton. The $\psi$ particles can live outside the soliton only in bound
states with vanishing $U(1)$ charge. The theory, however, presents another
spinor particle corresponding to the excitations of the Thirring field $\chi$,
which have zero $U(1)$ charge. However, according to Coleman's interpretation
of the sine-Gordon/Thirring equivalence, such excitations correspond to the
solitons themselves. Therefore, we can make an analogy with what is expected
to happen in QCD. The $\psi$ and $\chi$ particles are like the quarks and 
 hadrons respectively. The $U(1)$ charge is analogous to color in QCD, since
it is also confined. For the rigorous treatment of the dynamical flavor and color symmetries of the ATM model, as well as their relevant realizations, and the role played in a topological confinement mechanism, see a companion paper \ct{topological}. Then, our model can
be considered as a one dimensional bag model for QCD \ct{bla2, nowak}.

It could be interesting to compute the quantum corrections to the mass of the solitons associated to the $\vp$ field of the ATM model, for example, in the lines of Refs. \ct{dashen}. In our case, the quantum fluctuations of the spinor fields, of course, must be taken into account.

\appendix

\section{Appendix: Notations and Conventions}
\label{appa}

We use the following conventions in two dimensions. The metric tensor is $g_{\mu\nu}=\mbox{diag}(1, -1)$ and the antisymmetric tensor $\epsilon_{\mu\nu}$ is defined so that $\epsilon_{01}=-\epsilon^{01}=-1$. $\pa_{\pm}$ are derivatives w.r.t. to the light cone variables $x_{\pm}=t\pm x$. The gamma matrices are in the following representation:
\br
\gamma _{0}=-i\left( 
\begin{array}{cc}
0 & -1 \\ 
1 & 0
\end{array}
\right),\,\,\,\gamma _{1}=-i\left( 
\begin{array}{cc}
0 & 1 \\ 
1 & 0
\end{array}
\right) , \,\,\,\, \gamma _{5}=\gamma _{0}\gamma _{1} =\left( 
\begin{array}{cc}
1 & 0 \\ 
0 & -1
\end{array}
\right)
\er
satisfying anticommutation relations 
\br
\{\gamma _{\mu },\gamma _{\nu }\}=2g_{\mu \nu }{\bf 1,} 
\er
so the spinors $\psi$ and $\bar{\psi}$ are of the form
\br
\psi =\left( 
\begin{array}{c}
\psi _{R} \\ 
\psi _{L}
\end{array}
\right) ,\qquad \widetilde{\psi }=\left( 
\begin{array}{c}
\widetilde{\psi }_{R} \\ 
\widetilde{\psi }_{L}
\end{array}
\right) , \qquad \bar{\psi }=\left( 
\begin{array}{cc}
\widetilde{\psi }_{R}\,\, 
\widetilde{\psi }_{L}
\end{array}
\right) \gamma_{0}.
\er

\vspace{1cm}

\noindent{\bf Acknowledgements}

I would like to thank Professor L.A. Ferreira for his colaboration on parts of this work, and Professors G.M. Sotkov and A.H. Zimerman for valuable discussions. R. Bent\'{\i}n and C. Tello are also acknowledged for interesting conversations. I am grateful to Professor H.G. Valqui for introducing me to the study of non-linear phenomena and solitons. Thanks are also due to the organizers of the VII Hadron Physics 2000 for a very enjoyable Workshop. This work has been supported by FAPESP.

\end{document}